# Achieving Human Level Partial Credit Grading of Written Responses to Physics Conceptual Question using GPT-3.5 with Only Prompt Engineering

Zhongzhou Chen and Tong Wan

*Department of Physics, University of Central Florida, 4111 Libra Drive, Orlando, Florida, USA 32816*

Large language modules (LLMs) have great potential for auto-grading student written responses to physics problems due to their capacity to process and generate natural language. In this explorative study, we use a prompt engineering technique, which we name "scaffolded chain of thought (COT)", to instruct GPT-3.5 to grade student written responses to a physics conceptual question. Compared to common COT prompting, scaffolded COT prompts GPT-3.5 to explicitly compare student responses to a detailed, well-explained rubric before generating the grading outcome. We show that when compared to human raters, the grading accuracy of GPT-3.5 using scaffolded COT is 20% - 30% higher than conventional COT. The level of agreement between AI and human raters can reach 70% - 80%, comparable to the level between two human raters. This shows promise that an LLM-based AI grader can achieve human-level grading accuracy on a physics conceptual problem using prompt engineering techniques alone.

## I. INTRODUCTION

Generative AI (GenAI), especially large language models (LLMs) such as GPT and Llama, has been increasingly applied in science education [1], due to their ability to generate natural language output based on natural language input from a human user. Prior research has demonstrated LLMs' ability to solve physics [2] and math problems [3], provide personalized feedback to student written responses to a physics conceptual question [4], and grade student written responses to science questions [5].

In particular, LLMs' ability to "understand" students' written responses to a question and assign grades based on human-written rubric have a great potential to be widely applied in university physics and STEM classrooms, especially large enrollment courses. This would significantly reduce the grading loads of instructors and/or teaching assistants and enhance the quality of assessments. However, one of the well-known key drawbacks of LLMs is their tendency to "hallucinate", which means LLMs can generate outputs that are factually false or contextually implausible [6]. In grading student written responses, hallucination can result in LLM generating erroneous grading outcomes (including grades and justifications) that have a low level of agreement with human graders.

To reduce hallucination and increase the performance of LLMs, a few methods have been proposed in AI literature, including fine tuning [7], retrieval augmented generation (RAG) [8] few-shot learning [9], and prompt engineering [10]. Fine turning and RAG both require hundreds to thousands of pre-labeled data, which significantly hinders their applicability to grading, especially grading of new problems. Few-shot learning requires significantly less examples, but the outcome can be significantly impacted by the choice of examples. Prompt engineering, on the other hand, is a time- and cost-efficient method as it does not require examples or pre-labeled data.

A prompt is the natural language input by a human user to an LLM. Prompt engineering is the process of developing and refining a prompt to optimize the output by LLMs [10]. Many prompting techniques have been suggested to improve LLMs' ability for various tasks, such as Chain of thought (COT) prompting [11] and generated knowledge prompting [12]. A chain-of-thought prompt instructs an LLM to first generate a chain of intermediate reasoning steps before it generates the final answer. In some cases, LLM's performance can be significantly improved simply by adding "let's think step by step" at the end of the prompt. However, as is demonstrated in this study, the simple COT technique is still not sufficient to eliminate hallucinations during grading to a satisfactory level.

In this pilot study, we demonstrate that, by using a carefully engineered prompt, which we name "scaffolded COT", the accuracy of an AI grader can increase by 20% - 30% compared to simple COT prompting. The prompt with scaffolded COT is improved in two aspects. First, a detailed explanation of the rubric is provided together with the rubric. This is similar to explaining in detail each rubric item to a human rater. Second, the scaffolded COT prompt "forces" the LLM to first select the most relevant portion of the student answer, and then explicitly compare it to the rubric explanation before generating a grade. Using scaffolded COT, the level of agreement between an AI grader and human raters can reach 70% - 80%, which is comparable to the level between two human raters. This shows the potential that an LLM based AI grader can achieve human-level grading accuracy on a physics conceptual problem using prompt engineering techniques alone.

## II. METHODS

### A. Instructional context

The study was carried out in a large public research university in the south-eastern U.S. The target course was a calculus-based introductory physics course that focuses on Mechanics. The course was taught in a studio mode, which integrates lecture, recitation (or tutorial), and laboratory. The course enrollment was 99. On a midterm exam, students were given an opportunity to provide their explanations to their answers on a multiple-choice question. Students were told that in the case that they chose the wrong answer to the problem, they would be awarded partial credit according to their explanation of their reasoning.,

### B. Physics problem and grading rubric

The multiple-choice question used in this study concerns two swimmers sliding down frictionless water slides as shown in Fig. 1. The slides have identical height, but one is straight and the other is curved. Students were asked which swimmer, if either, will have a greater speed at the end of the slide. In a follow-up question, students were asked to explain their reasoning. They were prompted to specify the physics principle they used, why they think the principle can be applied, and the steps they used to reach the conclusion.

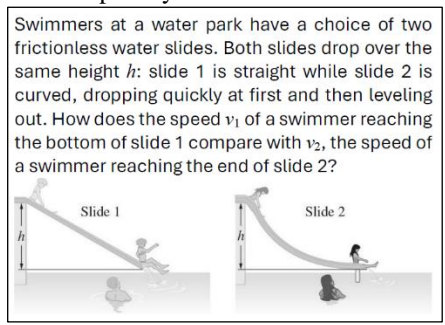

FIG. 1. The physics conceptual question used in GenAI grading.

The rubric we provided to the LLM has 3 items. The first item requires students to state that they used conservation of energy or work-energy theorem. The second item requires students to explain that because there is no non-conservative force doing work on the swimmer+earth system, the mechanical energy is conserved (or the work done by gravity causes the kinetic energy of the swimmer to change). The third item requires students to state that the potential energy is converted into kinetic energy of the system (or the work done by gravity is equal to the change in kinetic energy of the swimmer). Each rubric item has binary ratings, 1 or 0.

### C. Data collection

On the exam, 94 students provided explanations to their answers on the multiple-choice question. In this pilot study, we only included explanations from students who answered the multiple-choice question incorrectly (N = 40). Two human raters (i.e., the authors) independently graded the student explanations using the rubric described above. There were two reasons for choosing only the explanations associated with incorrect choices. First, from a practical perspective, only students with incorrect choices required partial credit grading, so the instructor initially only graded the explanation of the incorrect answer choices. Second, from a research perspective, the reasoning used by those who selected incorrect answers had a wider variety, ranging from completely incorrect to mostly correct with a minor mistake. On the other hand, explanations associated with correct answer choices were overall more uniform. Therefore, to test GenAI's ability to differentiate between different types of reasoning, we chose to focus on the reasoning for incorrect answers first.

### D. GenAI grading and prompt development

We used GPT-3.5 Turbo in completion mode, developed by OpenAI and accessed through the Microsoft Azure platform. The specific deployment of the LLM is conducted by university IT and all the data remained on university owned business level secure server. Access to and communication with the LLM was done using the LangChain AzureOpenAI python API.

The prompt given to GPT-3.5 Turbo consists of five components: (1) contextual information (e.g., introductory level physics course), (2) general grading instructions (e.g., binary ratings for each rubric item), (3) problem statement, (4) the rubric, and (5) specific grading requirements.

The prompt was developed and refined iteratively using five student explanations that were randomly selected. We developed and tested three versions of the prompt, which we name Naïve COT, Detailed-rubric COT, and Scaffolded COT. Each version has identical components (1) – (3), but different components (4) and (5), which are explained below.

Naïve COT: For component 4, the original rubric text is included with only minor modification. For component 5, the specific grading requirement states: "For each rubric item, first write step-by-step reasoning on why or why not the student explanation satisfies or contradicts the item, then conclude with a binary grade of either 1 or 0 for the rubric item."

Detailed-rubric COT: For component 4, each rubric item is accompanied by additional explanation text such as "student's explanation must explicitly contain 'frictionless' or 'smooth' or a similar phrase." The rubric items are also formatted slightly differently. For component 5, the grading requirement starts with the statement that "For each rubric item, first write a step-by-step reasoning that compares the student explanation to the rubric item and its contents."

Scaffolded COT: Component 4 is identical to that of Detailed-rubric COT. For component 5, the following text is provided.

```
# For each rubric item, write the grading statement
strictly following the order of the statements below:
## First, state one of the following two
    "For item <<item number>>, the rubric states that
    <<quote from the rubric item description>>. The
    most relevant parts in the student explanation are
    <<direct quote or quotes from student
    explanation>>."
    "For item <<item number>>, the rubric states that
    <<quote from the rubric item description>>. No part
    in the students' explanation is relevant to the
    rubric."
## then state one of the following:
    "the student explanation is similar to this part of
    the rubric description <<most similar part of the
    rubric>>,"
    "the student explanation and the rubric description
    are very different"
    "the student explanation and the rubric description
    are irrelevant"
## Finally, conclude with a binary score:
    "so the grade is 1"
    "so the grade is 0"
```

### E. Evaluation of grading accuracy

GPT-3.5 Turbo's grading accuracy is evaluated based on the level of agreement with human raters. The level of agreement (or disagreement) is quantified using three different metrics: percent agreement ($P_{agree}$), mean simple matching distance (SMD) [13,14], and quadratic weighted kappa (QWK) [15]. We also use the same metrics to quantify agreement between the two human raters, which can be used as a baseline for comparison.

$P_{agree}$ is the percentage of cases in which two raters agree in grading out of all three rubric items for all student responses. We calculated the $P_{agree}$ value between the AI grader and each human rater for each prompt version. To evaluate whether the three $P_{agree}$ values between the three prompt versions are significantly different, we used Cochran's Q test [16], an extension of the McNemar's test. A significant Q test result would indicate that fraction of complete agreement with human rater among the three prompt versions are not uniform.

SMD for two objects of $n$ binary attributes is defined as:
$$SMD = \frac{number\ of\ mismatched\ attributes}{total\ number\ of\ attributes}.$$

In our case, the SMD for a single response between the two raters (either human or AI) with three rubric items ($n = 3$) can take any of the four values: $0, \frac{1}{3}, \frac{2}{3}, 1$, with 0 representing perfect agreement and 1 representing complete disagreement. For each prompt version, we calculate SMD between the AI grader and one of the human raters for each student response, resulting in a set of 40 SMD values for each prompt version against one human rater. To compare the distributions of SMD values resulting from the three different prompt versions, we use the Friedman's test [17], an extension of the Wilcoxon signed-rank test. The Friedman's test in this case examines whether one of the sets of SMD values are consistently larger or smaller compared to the other sets. This is a non-parametric test that does not rely on the distribution of the underlying data set.

QWK, sometime referred to as the weighted Cohen's Kappa, measures the similarity in total score between two raters. Cohen's Kappa is used for categorical variables, while QWK is used for ordinal variables (i.e., variables that can be sorted or ranked). The total score assigned to a student response can be 0, 1, 2, or 3, and thus it is an ordinal variable. QWK is widely reported in auto-grading literature and reflects the agreement in total score that is assigned to a student response. The interpretation of the QWK result is the same as that of Cohen's Kappa. However, it must be noted that since QWK is generated based on the total score rather than the score of each rubric, it may not reflect the level of agreement when two raters assign the same total score but differ on individual rubric items.

### III. RESULTS

Table I shows the quantified agreement or disagreement between the AI grader and a human rater measured by the three metrics. The agreement or disagreement between the two human raters is also shown for comparison. Results of all three metrics show the same trend: Naïve COT prompting is somewhat less accurate than Detailed-rubric COT, and Scaffolded COT grading is significantly more accurate than the other two prompts.

It is worth noting that the level of agreement measured by $P_{agree}$ between Scaffolded COT and the human graders are comparable to the level of agreement between the human raters, between 70% - 80%, which is significantly higher than that of the other two prompts. The same can be said regarding the mean SMD metric for measuring disagreement.

Statistical tests also confirmed that the agreement (or disagreement) with human rater 1 was improved by refining the prompt ($p = 0.004$ for Cochran's Q and $p = 0.031$ for Friedman's). That is, the agreement for the Scaffolded COT with human rater 1 was higher than the agreement for at least one of the other two prompt versions. When compared with human rater 2, the result from the Cochran's Q test ($p = 0.050$) is very close to the significance threshold, while the result from Friedman's test ($p = 0.062$) shows that the improvement is not significant. Nonetheless, the QWK results show that the Scaffolded COT has an improved level of agreement with both human raters (from substantial to almost perfect for human rater 1, and from moderate to substantial for rater 2).

Another interesting observation is that the Scaffolded COT results seem to agree with human rater 1 more than human rater 2. It may be because human rater 1 developed the detailed rubric explanations used in both Detailed-rubric COT and Scaffolded COT.

TABLE I. Agreement or disagreement between the AI grader, for each prompt version, and each human rater measured by three metrics. The agreement between the two human raters as a baseline is also shown.

|  | Percent Agreement | | Mean Simple Matching Distance | | Quadratic Weighted Kappa[†] | |
| --- | --- | --- | --- | --- | --- | --- |
|  | Rater 1[**] | Rater 2[‡] | Rater 1[*] | Rater 2 | Rater 1 | Rater 2 |
| Naïve COT vs. Human Rater | 55% | 50% | 0.21 | 0.22 | 0.68 | 0.58 |
| Detailed-rubric COT vs Human Rater | 60% | 60% | 0.21 | 0.22 | 0.66 | 0.57 |
| Scaffolded COT vs. Human Rater | 83% | 70% | 0.08 | 0.11 | 0.84 | 0.66 |
| Human Rater 1 vs. Human Rater 2 | 75% | | 0.09 | | 0.73 | |

[†]0.41-0.60, moderate; 0.61-0.80, substantial; 0.81-1, almost perfect. [*]$p < 0.05$. [**]$p < 0.01$. [‡]$p = 0.05$. Note that the statistical significance pertains to the differences in human agreement level between the three prompt versions.

### IV. CONCLUSIONS AND FUTURE DIRECTIONS

In this study, we demonstrated that a significant increase in grading accuracy of students' written responses to a conceptual question can be achieved solely through prompt engineering, without the need for more sophisticated techniques such as few-shot learning, fine-tuning, or RAG. For the one problem tested in this study, the accuracy of the grading using the Scaffolded COT prompt is comparable to human raters. The results could help develop an AI-based grading system with significantly lower cost using better designed prompts.

Scaffolded COT can be seen as a stronger form of COT that integrates some of the features of generated knowledge prompting, which prompts LLMs to first generate relevant and useful information before the final answer. We hypothesize that the superior performance of this prompt style results from two factors. First, the scaffold structure forces the LLM to consistently generate the reasoning prior to making the conclusion, whereas Naïve COT prompt did not consistently produce this behavior. Second, Scaffolded COT strongly forces the LLM to generate reasoning based on an explicit comparison between a student answer to the detailed rubric explanations, which prevents the LLM from making up reasons that superficially seems legitimate.

However, it must also be pointed out that since LLMs are stochastic systems, the study needs to be re-conducted in the future multiple times to test whether the results are reproducible. Moreover, this explorative study only tested the grading of responses from students who chose the wrong answers, of which more than a half received 0 for all three rubric items. In fact, the authors have recently experimented with grading the complete dataset of 99 responses, using both GPT-3.5 Turbo in chat mode, as well as GPT-4o in chat mode. Preliminary data found that the performance of Scaffolded COT prompt is not stable on multiple runs, but GPT-4o is able to reliably deliver the same level of performance (~75% agreement with human raters) using Detailed-Rubric COT. This updated result will be reported in a future publication.

Finally, in the current study the first author designed the prompt and also graded students' responses. The design of prompts, especially the detailed rubric in the prompt, could have been biased by having seen the student responses. In future studies it will be more desirable if grading and prompt designing could be conducted by two different people.

---

[1] E. Kasneci et al., *ChatGPT for Good? On Opportunities and Challenges of Large Language Models for Education*, Learn Individ Differ **103**, (2023).

[2] G. Polverini and B. Gregorcic, *How Understanding Large Language Models Can Inform the Use of ChatGPT in Physics Education*, Eur J Phys (2023).

[3] M. Zong and B. Krishnamachari, *Solving Math Word Problems Concerning Systems of Equations with GPT-3*, in *The Thirty-Seventh AAAI Conference on Artificial Intelligence (AAAI-23)* (2023).

[4] T. Wan and Z. Chen, *Exploring Generative AI Assisted Feedback Writing for Students' Written Responses to a Physics Conceptual Question with Prompt Engineering and Few-Shot Learning*, Phys Rev Phys Educ Res **20**, 010152 (2024).

[5] E. Latif and X. Zhai, *Fine-Tuning ChatGPT for Automatic Scoring*, Computers and Education: Artificial Intelligence **6**, 100210 (2024).

[6] M. Lee, *A Mathematical Investigation of Hallucination and Creativity in GPT Models*, Mathematics **11**, (2023).

[7] E. Radiya-Dixit and X. Wang, *How Fine Can Fine-Tuning Be? Learning Efficient Language Models*, in *23rdInternational Conference on Artificial In- Telligence and Statistics (AISTATS)* (PMLR: Volume 108., Palermo, Italy, 2020).

[8] P. Lewis et al., *Retrieval-Augmented Generation for Knowledge-Intensive NLP Tasks*, Adv Neural Inf Process Syst **33**, 9459 (2020).

[9] Y. Wang, Q. Yao, J. T. Kwok, and L. M. Ni, *Generalizing from a Few Examples: A Survey on Few-Shot Learning*, ACM Comput Surv **53**, (2020).

[10] P. Liu, W. Yuan, J. Fu, Z. Jiang, H. Hayashi, and G. Neubig, *Pre-Train, Prompt, and Predict: A Systematic Survey of Prompting Methods in Natural Language Processing*, ArXiv:2107.13586v1 (2021).

[11] J. Wei, X. Wang, D. Schuurmans, M. Bosma, B. Ichter, F. Xia, E. Chi, Q. Le, and D. Zhou, *Chain-of-Thought Prompting Elicits Reasoning in Large Language Models*, ArXiv:2201.11903v6 (2023).

[12] J. Liu, A. Liu, X. Lu, S. Welleck, P. West, R. Le Bras, Y. Choi, and H. Hajishirzi, *Generated Knowledge Prompting for Commonsense Reasoning*, ArXiv:2110.08387v3 (2021).

[13] H. Prasetyo and A. Purwarianti, *Comparison of Distance and Dissimilarity Measures for Clustering Data with Mix Attribute Types*, 2014 1st International Conference on Information Technology, Computer, and Electrical Engineering: Green Technology and Its Applications for a Better Future, ICITACEE 2014 - Proceedings 276 (2015).

[14] G. Gan, C. (Professor) Ma, and J. Wu, *Data Clustering: Theory, Algorithms, and Applications*, 466 (2007).

[15] S. Vanbelle, *A New Interpretation of the Weighted Kappa Coefficients*, Psychometrika **81**, 399 (2016).

[16] W. G. Cochran, *The Comparison of Percentages in Matched Samples*, Biometrika **37**, 256 (1950).

[17] M. Friedman, *The Use of Ranks to Avoid the Assumption of Normality Implicit in the Analysis of Variance*, J Am Stat Assoc **32**, 675 (1937).